\documentstyle[twocolumn,aps,psfig]{revtex}

\begin{document}
\draft
\flushbottom
\twocolumn[
\hsize\textwidth\columnwidth\hsize\csname @twocolumnfalse\endcsname

\title{ Kaluza-Klein theory of electron-plasmon interaction in a thin cylindrical wire}
\author{Igor I. Smolyaninov }
\address{ Department of Electrical and Computer Engineering \\
University of Maryland, College Park,\\
MD 20742}
\date{\today}
\maketitle
\tightenlines
\widetext
\advance\leftskip by 57pt
\advance\rightskip by 57pt

\begin{abstract}
Electron-plasmon interaction in a thin cylindrical wire is described in terms of four-dimensional Kaluza-Klein theory in which angular coordinate of the metal cylinder is considered to be one of the two compactified coordinates. In this model electron-electron, electron-plasmon and nonlinear optical plasmon-plasmon interactions are unified via introduction of effective superchiral charges and fields. Such an approach has important consequences for description of electromagnetic and transport properties of mesoscopic metallic wires and rings at low temperatures, such as recent observation of density-dependent spin polarization in quantum wires. This theory may be consistently reformulated using normal language of classical physics as nonlinear electrodynamics of a rotating medium.   

\end{abstract}

]
\narrowtext

\tightenlines

Very recently a novel description of nonlinear optics of cylindrical surface plasmons, which may be relevant in single-photon tunneling effect \cite{1} has been introduced. Cylindrical surface plasmon (CSP) mode coupling in a nanowire or a nanochannel has been described in terms of a three-dimensional Kaluza-Klein theory in which the angular coordinate of the metal cylinder is considered as a compactified dimension \cite{2}. Higher ($n>0$) CSP modes have been shown to posses quantized effective chiral charges proportional to their angular momenta $n$. In a nanowire these slow moving charges exhibit long-range interaction via exchange of fast massless CSPs with zero angular momentum. These zero angular momentum CSPs may be considered as quanta of the gyration field (the field of gyration vector $\vec{g}$, which relates the $\vec{D}$ and $\vec{E}$ fields in an optically active medium). This mode-coupling theory may be used in description of nonlinear optics of cylindrical nanowires and nanochannels (for example, in single-photon tunneling effect), and may be further extended to describe mode interaction in other chiral optical waveguides. Moreover, after the fact, this theory may be consistently re-formulated using normal language of nonlinear Maxwell equations \cite{2}, although this fact does not diminish the value of the insights gained by the Kaluza-Klein consideration. Generally, a nonlinear system with interaction is very difficult to analyze, so the hints obtained from consideration of similar well-understood systems are very useful.  

In this Letter I further extend the Kaluza-Klein analogy to gain more understanding of the mesoscopic physics of thin cylindrical metal wires. I am going to show that electron states with nonzero angular momenta exhibit similar long range chiral interaction. Some similarity between the chiral charges of the cylindrical surface plasmons and the electron states with nonzero angular momenta has been already noted in \cite{2}, and its relevance for mesoscopic conductance of thin cylindrical wires and carbon nanotubes has been conjectured. This similarity stems from the way in which the chiral charges have been introduced. Similar to the original five-dimensional Kaluza-Klein theory \cite{3}, nonlinear optics of cylindrical surface plasmons has been considered as if it occurs in a space-time which has a small "compactified" dimension. Since CSPs may be considered as if they live in a "three-dimensional space-time" with two spatial dimensions being a very long z-dimension along the axis of the cylinder, and a small "compactified" $\phi $-dimension along the circumference of the cylinder, the theory of CSP mode propagation and interaction may be formulated as a three-dimensional Kaluza-Klein theory. On the other hand, in the original five-dimensional theory the normal electric charges themselves are introduced as chiral (nonzero angular momentum) modes of a massless quantum field, which is quantized over the cyclic compactified fifth dimension. Thus, similarity of the effective CSP chiral charges and the normal electric charges becomes evident. 

The logical continuation of these ideas is to consider both the chiral charges of the CSPs and the normal electric charges in a unified way, in which electron-electron, electron-plasmon, and nonlinear optical plasmon-plasmon interaction in a thin metal wire is described as if it happens in a four-dimensional space-time, which besides the extended z-coordinate has two compactified cyclic spatial dimensions: the $\phi $-dimension along the circumference of the metal cylindrical wire, and the cyclic fifth dimension of the original Kaluza-Klein theory. Similar to high energy physics where the search for the underlying universal symmetries between different particles and fields has been a very fruitful approach, we may expect that such a unified description may bring about new deep insights into the mesoscopic physics. Such insights are highly necessary taking into account numerous current contradictions between mesoscopic theories and experiment \cite{4,5}. At least some of these contradictions may be explained by previously unaccounted intrinsic electron-electron interactions \cite{4}. Electron-electron interaction via exchange of cylindrical surface plasmons may be one of such interactions. 

In modern Kaluza-Klein theories the extra N-4 space-time dimensions are considered to be compact and small (with characteristic size on the order of the Planck length). The symmetries of this internal space are chosen to be the gauge symmetries of some gauge theory \cite{6}, so a unified theory would contain gravity together with the other observed fields. In the original form of the theory a five-dimensional space-time was introduced where the four dimensions $x^1, ..., x^4$ were identified with the observed space-time. The associated 10 components of the metric tensor $g_{\alpha \beta }$ were used to describe gravity. After a compactified fifth dimension $x^5$ with a small circumference $L$ was added, the extra four metric components $g_{\alpha 5}$ connecting $x^5$ to $x^1, ..., x^4$ gave four extra degrees of freedom which were interpreted as the electromagnetic potential (here we use the following convention for greek and latin indices: $\alpha = 1, ..., 4$; $i = 1, ..., 5$). An additional scalar field $g_{55}$ or dilaton may be either set to a constant, or allowed to vary. 

When a quantum field $\psi $ coupled to this metric via an equation

\begin{equation} 
\Box_5 \psi +a\psi =0
\end{equation}

is considered, where $\Box_5$ is the covariant five-dimensional d'Alembert operator, the solutions for the field $\psi $ must be periodic in the $x^5$ coordinate. This leads to the appearance of an infinite "tower" of solutions with quantized $x^5$-component of the momentum: 

\begin{equation}
q^5_n=2\pi n/L 
\end{equation}

where n is an integer. In our four-dimensional space-time on a large scale such solutions with $n\neq 0$ interact with the electromagnetic potential $g_{\alpha 5}$ as charged particles with an electric charge $e_n$ and mass $m_n$:

\begin{equation}
e_n = \hbar q_n(16\pi G)^{1/2}/c
\end{equation}

\begin{equation}
m_n = \hbar (q_n^2-a)^{1/2}/c
\end{equation}

where $G$ is the gravitational constant (see for example the derivation in \cite{7}). It is well-known that Maxwell equations in a general curved space-time background $g_{ik}(x,t)$ are equivalent to the macroscopic Maxwell equations in the presence of matter background with some nontrivial electric and magnetic permeability tensors $\epsilon _{ik}(x,t)$ and $\mu _{ik}(x,t)$ \cite{8}. Thus, strong similarity between the results of the three-dimensional Kaluza-Klein theory and the solutions of nonlinear Maxwell equations in some quasi-three-dimensional cylindrical waveguide geometries is natural: in both cases we consider a massless field in a geometry which has a compactified cyclic $\phi $-dimension in the presence of nontrivial space-time curvature or permeability tensors, respectively. This similarity is especially strong in the case of surface plasmon waveguides \cite{2}, since surface plasmons live in (almost) three-dimensional space-times on the metal interfaces. Metal interfaces are nonlinear and discriminate between left and right circular polarizations due to magnetic field induced optical activity. Thus, higher ($n>0$) CSP modes posses quantized effective chiral charges proportional to their angular momenta $n$, and exhibit long-range interaction via exchange of fast massless zero angular momentum CSPs. 

Let us now follow this general Kaluza-Klein recipe, and try and build a unified theory of electron-plasmon interaction in a thin metal wire by considering a four-dimensional space-time, which besides the extended z-coordinate has two compactified cyclic spatial dimensions: the $\phi $-dimension along the circumference of the metal cylindrical wire, and the cyclic fifth $\theta $-dimension of the original Kaluza-Klein theory. As has been shown in \cite{2}, the effects of the radial coordinate can be included in the factor $a$ of equation (1), so we will neglect the radial coordinate of the real physical wire for the sake of simplicity. We will also assume temperature to be very low, so that interactions of only a few electrically charged quaziparticles can be considered, while the rest of the electrons in the metal are taken into account via electric and magnetic permeability tensors $\epsilon _{ik}$ and $\mu _{ik}$ of the metal, which in turn are taken into account via the space-time metric $g_{ik}$. Thus, the effective metric can be written as

\begin{eqnarray}
ds^2 = c^2dt^2 - dz^2 - R^2d\phi ^2 - r^2d\theta ^2+ 2g_{02}cdtd\phi + 
\nonumber \\
2g_{03}cdtd\theta + 2g_{12}dzd\phi + 2g_{13}dzd\theta + 2g_{23}d\theta d\phi
\end{eqnarray} 

where $R$ and $r$ are the radii of the cylindrical wire and the compactified original fifth Kaluza-Klein dimension, respectively. Here we are not interested in possible z-dependence of $R$ and consider all $g_{0i}$ and $g_{1i}$ components to be independent of $\phi $ and $\theta $. 
Equation (1) with $a=0$ for a quantum massless scalar field $\psi $ in this metric should be written as 

\begin{equation}
\frac{\partial }{\partial x^i}(g^{ik}\frac{\partial \psi }{\partial x^k})=0,
\end{equation}

where $i,k=0, ... , 3$. The field $\psi $ is considered to be scalar in order to make the consideration as simple as possible (for a vector field each component of the field will need to satisfy equation (6)). We will search for the solutions in the usual form as
$\psi = \Psi (x^{\alpha })e^{iq_2\phi }e^{iq_3\theta }$, where $\alpha =0,1$, and periodicity in $\phi $ and $\theta $ requires $q_2=n_2$ and $q_3=n_3$ ($n_2$ and $n_3$ are integer). As a result, we obtain

\begin{eqnarray}
(\frac{\partial ^2}{c^2\partial t^2}-\frac{\partial ^2}{\partial z^2})\psi -(q_2^2g^{22}+q_3^2g^{33}+2q_2q_3g^{23})\psi+ \nonumber \\ 
2iq_2g^{\alpha 2}\frac{\partial \psi}{\partial x^{\alpha }}+2iq_3g^{\alpha 3}\frac{\partial \psi}{\partial x^{\alpha }}+iq_2 (\frac{\partial g^{\alpha 2}}{\partial x^{\alpha }})\psi+iq_3 (\frac{\partial g^{\alpha 3}}{\partial x^{\alpha }})\psi = 0
\end{eqnarray} 

This is a two-dimensional Klein-Gordon equation describing a two-component quantized superchiral charge $(q_2;q_3)$ in the presence of external $g^{\alpha 2}(t,z)$ and $g^{\alpha 3}(t,z)$ vector fields (the name superchiral is chosen to emphasize the unified description of electrons and plasmons). The $q_3$ component of the superchiral charge and the $g^{\alpha 3}$ field correspond to the quantized electric charge $e \sim q_3$ and the electromagnetic field, respectively, while the $q_2$ component of the charge and the $g^{\alpha 2}$ field correspond to the chiral charge and the gyration field described in \cite{2}. Thus, we have obtained a unified symmetric description of both types of charges and their interactions. For example, it is easy to see that the term $2q_2q_3g^{23}\psi $ in equation (7) corresponds to the orbital magnetic moment of the electric charge $q_3$, which interacts with the axial magnetic field described by $g^{23}$. At the same time, equation (7) indicates that the axial magnetic field is not the only field acting on the electric charges which posses nonzero orbital momenta. According to (7), such states respond to the gyration field $g^{\alpha 2}$ in the same way as the CSPs with nonzero angular momenta ($n>0$). In other words, a rotating electric charge has an effective chiral charge, and it exhibits long-range interactions with other rotating electric charges and $n>0$ CSPs via exchange of fast massless CSPs with zero angular momentum.     

After we have understood the general symmetry and structure of the electron-plasmon interaction, we can reformulate the above description using normal language of nonlinear Maxwell and Schrodinger equations. In our consideration I will follow the Landau and Lifshitz \cite{9} way of introducing optical activity (gyration) tensor in the macroscopic Maxwell equations, and for simplicity use the following equation valid in isotropic or cubic-symmetry materials:

\begin{equation}  
\vec{D}=\epsilon \vec{E}+i\vec{E}\times \vec{g} ,
\end{equation}

where $\vec{g}$ is called the gyration vector. If the medium exhibits magneto-
optical effect, and does not exhibit natural optical activity, $\vec{g}$ is 
proportional to the magnetic field $\vec{H}$:

\begin{equation}  
\vec{g}=f\vec{H} ,
\end{equation}

where the constant $f$ may be either positive or negative. For metals in the 
Drude model at $\omega >>eH/mc$

\begin{equation}  
f(\omega )= -\frac{4\pi Ne^3}{cm^2\omega ^3}=-\frac{e\omega _p^2}{mc\omega ^3} ,
\end{equation}

where $\omega _p$ is the plasma frequency and $m$ is the electron mass \cite{9}.

The derivation of chiral interaction of the CSPs with each other from the nonlinear Maxwell equations has been presented in \cite{2}, while general properties of the cylindrical surface plasmons are described in \cite{10}. It has been shown in \cite{2} that the field of zero-angular-momentum CSPs corresponds to the gyration field $g^{\alpha 2}$ in (7) due to magneto-optical effect (eq. (9) and (10)) in the cylindrical metal wire. Let us start from the effective Poisson equation for the chiral charges and the gyration field in the cylindrical wire geometry in the form \cite{2}

\begin{equation}  
\vec{\nabla}\times (\vec{\nabla }\times \vec{B_0}-\frac{4\pi f\omega _n }{c^2}\vec{S_n})=0 
\end{equation}

where $\vec{B_0}$ is the magnetic field of the zero-angular-momentum CSP (gyration) field, and $\omega _n$ and $\vec{S_n}$ are the frequency and the Pointing vector of the n-th CSP mode, which acts as a source of the gyration field. By looking at the expression under the parenthesis in equation (11) it is clear that the electric current $\vec{j}$ should act as an additional source of $\vec{B_0}$, so that the effective Poisson equation takes the form

\begin{equation}  
\Delta \vec{B_0}=\frac{4\pi f\omega _n}{c^2}\vec{\nabla }\times 
\vec{S_n}+\frac{4\pi }{c}\vec{\nabla }\times \vec{j} 
\end{equation}

Thus, both the chiral charges of the CSPs and the rotating electric charges enter this equation in a symmetric way, and act as the sources of the gyration field. 

As a second step, let us consider how the field of a zero-angular-momentum CSP (gyration field) acts on a rotating electric charge. The $\omega (k)$ of the $n=0$ CSP mode goes down to $\omega =0$ approaching the light line $\omega =kc/\epsilon ^{1/2}$ from the right, as $k\rightarrow 0$. The field of this mode has only the following nonzero components: $E_r$, $E_z$, and $H_{\phi }$ \cite{10}, and may be described in cylindrical coordinates $(r,\phi ,z)$ by the vector potential of the form $\vec{A}=(A_r, 0, A_z)$, so that $E_r=\partial A_r/\partial t$, $E_z=\partial A_z/\partial t$, and $H_{\phi }=\partial A_r/\partial z-\partial A_z/\partial r$. The linear Shrodinger equation with the vector potential of this form can be used in evaluation of CSPs effect on the phase of the electron wave function. Unfortunately, the chiral interaction of interest is due to nonlinear optical effects, and arises as higher order terms in QED of a nonlinear optical medium. However, a good physical picture of the chiral interaction may be obtained using the analogy between the Maxwell equations in a given gravitational field \cite{11}, and the Maxwell equations in a chiral medium. Relationship between the $\vec{D}$ and $\vec{E}$ fields in a gravitational field with the metric $g_{ik}$ can be written in the form

\begin{equation}  
\vec{D}=\frac{\vec{E}}{h^{1/2}}+\vec{H}\times \vec{g} ,
\end{equation}

where $h=g_{00}$ and $g_{\alpha }=-g_{0\alpha }/g_{00}$. For an electromagnetic wave this relationship is very similar to equation (8) if we identify 
$\epsilon $ as $1/h^{1/2}$, and take into account different conventions for the use of imaginary numbers in \cite{9} and \cite{11}. Thus, the space-time itself is a chiral optical medium if $\vec{g}\neq 0$. Such a space-time region can be described locally as a rotating coordinate frame with an angular velocity \cite{11}

\begin{equation}  
\vec{\Omega }=\frac{ch^{1/2}}{2}\vec{\nabla }\times \vec{g} 
\end{equation}
 
Similar vector $\Omega $ field can be defined for any chiral medium regardless of the nature of its optical activity (natural or magnetic field induced).
It is clear that in the presence of $\vec{\Omega }\neq 0$ (as is the case for zero-momentum CSP propagating along the metal wire, due to the $H_{\phi }$ field of the CSP and eq.(9)) any particle with nonzero angular momentum (CSP or electron) acquires additional energy $\vec{L}\vec{\Omega }$, where $\vec{L}$ is the angular momemntum. This is the physical meaning of the terms in equation (7) describing the chiral interaction of the angular momenta with the gyration field.

As has been shown in \cite{2}, in a thin metal wire the chiral interaction via exchange of zero-angular-momentum CSPs becomes quite noticeable and long-range. Unlike Coulomb interaction, which is screened by the presence of other free electrons, chiral interaction does not experience much screening. Moreover, because of one-dimensional nature of this interaction, it does not depend on the distance between the chiral charges unless the CSP decay (finite free propagation length) is taken into account. All these factors may make the chiral electron-electron, electron-plasmon and plasmon-plasmon interactions an important mechanism in mesoscopic transport phenomena.     

A number of experiments to check the importance of chiral interactions may be suggested. For example, the CSP spectrum of a cylindrical metal wire may be changed by periodic modulation of the shape of the wire, and the effects of this change on different mesoscopic properties may be studied. Comparison of transport properties of wires with different cross-sections may also be performed. Breaking of the cylindrical symmetry of the wire may be described as an external chiral field, which affects the chiral electron-plasmon interaction, but does not remove it completely. An external chiral field may also be created by coating the wire with a layer of chiral optical material or external illumination with circular-polarized light. In all these cases the chiral electron energy level splitting may be studied. If the energy level splitting is high enough, one can start to talk about para- or dia-chiral response of the samples. These are just a few of possible experimental checks and developments of the ideas described in this paper.

I should also mention that density-dependent spin polarization recently observed in zero magnetic field in ultra-low-disorder quantum wires \cite{12} may be explained as a manifestation of chiral energy level splitting. According to the estimate below, electric current flowing through the quantum wire during measurements may be just high enough to create substantial gyration field $\vec{g}=f\vec{H}$ circulating around the wire, and hence, substantial 
$\vec{\Omega}\sim \vec{\nabla }\times \vec{g}$ field directed along the wire. This will induce energy level splitting $\Delta E=\hbar \Omega $ between the spin up and down electron states. Let us estimate the order of magnitude of this effect. Taking into account equations (9) and (14), the chiral splitting can be written as

\begin{equation}
\Delta E \sim \frac{2\hbar f I}{r^2} ,  
\end{equation}

where $I$ is the current through the quantum wire (according to the data presented in \cite{12}, $I\sim 0.25 mA$), and $r$ is its characteristic radius, which I assume to be around $r\sim 10 nm$. Taking as an estimate the value of $f$ at $\omega =\pi c/l$ from equation (10), where $l=1.0 \mu m$ is the sample length in \cite{12}, we obtain $\Delta E \sim 5\times 10^{-4} eV=5 K$. Although more precise theoretical treatment of this effect will be necessary in the future, the above estimate indicates that the chiral energy level splitting is an observable effect, and it may in fact already been observed. In addition, the fact that the $f$ constant of the magneto-optical effect is proportional to the electron density in the wire corroborate phenomenological picture of \cite{12} indicating an electron density-dependent spin energy gap.

In conclusion, electron-plasmon interaction in a thin cylindrical wire has been described in terms of four-dimensional Kaluza-Klein theory in which angular coordinate of the metal cylinder is considered to be one of the two compactified coordinates. In this model electron-electron, electron-plasmon and nonlinear optical plasmon-plasmon interactions are unified via introduction of effective superchiral charges and fields. Such an approach has important consequences for description of electromagnetic and transport properties of mesoscopic metallic wires and rings at low temperatures, and may be consistently reformulated using normal language of classical physics.

\end{document}